\pdfoutput=1

\documentclass[11pt]{article}

\usepackage[final]{acl}

\usepackage{times}
\usepackage{latexsym}
\usepackage{hyperref}
\usepackage{subcaption}
\usepackage{adjustbox}
\usepackage{multirow}
\usepackage{tikz}
\usepackage[normalem]{ulem}
\useunder{\uline}{\ul}{}

\usepackage[T1]{fontenc}

\usepackage[utf8]{inputenc}

\usepackage{microtype}

\usepackage{inconsolata}

\usepackage{graphicx}

\title{CatMemo at the FinLLM Challenge Task: Fine-Tuning Large Language Models using Data Fusion in Financial Applications}

\author{Yupeng Cao{*},~ 
Zhiyuan Yao{*},~
Zhi Chen{*},~
Zhiyang Deng{*}\\
{*}Equal Contribution\\
Stevens Institute of Technology, Hoboken, NJ\\
\texttt{\{ycao33,zyao9,zchen100,zdeng10\}@stevens.edu}
}

\begin{document}

{\makeatletter
}
\maketitle

\begin{abstract}
The integration of Large Language Models (LLMs) into financial analysis has garnered significant attention in the NLP community. This paper presents our solution to IJCAI-2024 FinLLM challenge, investigating the capabilities of LLMs within three critical areas of financial tasks: financial classification, financial text summarization, and single stock trading. We adopted Llama3-8B and Mistral-7B as base models, fine-tuning them through Parameter Efficient Fine-Tuning (PEFT) and Low-Rank Adaptation (LoRA) approaches. To enhance model performance, we combine datasets from task 1 and task 2 for data fusion. Our approach aims to tackle these diverse tasks in a comprehensive and integrated manner, showcasing LLMs' capacity to address diverse and complex financial tasks with improved accuracy and decision-making capabilities.

\end{abstract}

\section{Introduction}
In recent years, FinTech research has increasingly focused on using textual information to aid investment decisions by analyzing various financial textual data~\cite{allen2021survey}. However, the complexity of financial documents makes it difficult to classify and summarize market information. Additionally, the intricate and volatile nature of financial markets poses significant challenges for making informed, sequential investment decisions. To address these challenges, advanced natural language processing techniques and models are necessary to process and interpret vast amounts of financial data accurately~\cite{fisher2016natural}. Lately, Large Language Models (LLMs) have demonstrated impressive capabilities in the field of finance \cite{bubeck2023sparks, li2023large}. These models excel in understanding and generating human-like text, making them ideal candidates for tackling complex financial tasks. 

Although LLMs demonstrate significant promise in the financial sector, their efficacy in specific financial tasks requires deeper investigation. The FinLLM challenge @ IJCAI-2024 initiative, as introduced in~\citet{xie2024finben}, seeks to investigate the potential of LLMs in analyzing financial documents and enhancing decision-making processes. By leveraging the power of LLMs, the initiative aims to improve the accuracy and efficiency of financial information processing, ultimately aiding in improved investment strategies and a better market understanding.

This paper describes our technical solution for three diverse tasks provided by the  FinLLM challenge: financial classification~\cite{sy2023fine}, text summarization~\cite{zhou2021trade}, and single stock trading \cite{yu2024finmem}. The classification task involves distinguishing between claims and premises in financial texts, the summarization task aims to distill extensive financial narratives into succinct summaries, and the trading task focuses on formulating predictive trading decisions based on algorithmic insights.

The core idea of our solution is to fine-tune pre-trained LLMs using PEFT \cite{peft} and LoRA \cite{hu2021lora} techniques, leveraging data fusion strategy on the provided datasets from task 1 \& 2 in the FinLLM challenge. Specifically, we select Llama3-8B~\cite{llama3modelcard} and Mistral-7B~\cite{jiang2023mistral} as the pre-trained base models due to their large number of parameters, which enable them to capture complex patterns and nuances in financial text data—essential for the three tasks in the challenge. Additionally, these models are pre-trained on vast and diverse datasets, providing a broad understanding of language that can be fine-tuned for financial domains, enhancing their versatility and adaptability to specific financial tasks. Furthermore, both models support PEFT and LoRA techniques, allowing efficient and effective specialization for the financial domain, even with limited labeled data.

Our extensive experiments conducted on the three shared tasks have yielded significant findings: 1) Mistral-7B outperforms Llama3-8B in terms of both overall performance and its ability to generate well-structured outputs; 2) the fine-tuned model by using the fused data, showed enhanced results on Task 1 and Task 2; 3) however, this fine-tuned model did not demonstrate improvement in the more complex single-stock trading task (Task 3). For this, we do a more detailed analysis of the results in Section 4.


\section{Shared Task Description}
The FinLLM challenge consists of three shared tasks: financial classification (task 1), text summarization (task 2), and single stock trading (task 3). Datasets description can be found in: \url{https://huggingface.co/datasets/TheFinAI/flare-finarg-ecc-auc_test} and \url{https://huggingface.co/datasets/TheFinAI/flare-edtsum_test}.

\noindent
\textbf{Task 1} in the FinLLM challenge focuses on the \textbf{financial classification}, specifically categorizing sentences within financial documents as either claims or premises. A claim is a statement that asserts a point of view or opinion, while a premise provides the supporting information or evidence for that claim. This task is fundamental for understanding and analyzing financial narratives, as it helps in structuring the information into coherent arguments, which is essential for various downstream applications such as sentiment analysis, risk assessment, and investment decision-making. The evaluation metric for Task 1 is the \textbf{F1 score}, which provides a balanced measure of the model's precision and recall.\\~\\
\textbf{Task 2} in the FinLLM challenge focuses on \textbf{financial texts summarization}. The objective is to condense lengthy financial documents into concise summaries that capture the essential information and key insights while omitting redundant or less important details. This task is crucial for enabling quick and effective information processing, allowing stakeholders to make informed decisions without wading through extensive reports. Task 2 utilizes three metrics, namely ROUGE (1, 2, and L) and BERTScore, to evaluate generated summaries in terms of relevance, with the \textbf{ROUGE-1 score} serving as the final ranking metric.\\~\\
\textbf{Task 3} in the FinLLM challenge focuses on the application of LLMs to \textbf{single stock trading}, aiming to make informed and predictive trading decisions. The primary goal of this task is to develop a model that can analyze various financial texts and other relevant data to predict the future price movements of a single stock and make trading decisions based on these predictions. The evaluation metric includes Sharpe Ratio (SR), Cumulative Return (CR), Daily (DV) and Annualized Volatility (AV), and Maximum Drawdown (MD), with the \textbf{Sharpe Ratio (SR)} used as the final ranking metric.

\section{Proposed Method}
\begin{figure}[ht]
\centering
\includegraphics[width=0.95\linewidth]{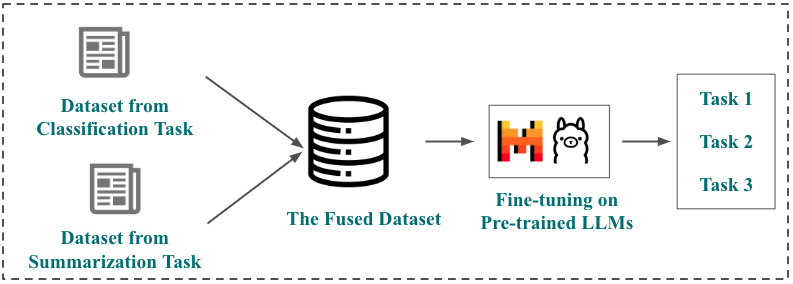}
\caption{Schematic of proposed fine-tuning method.}
\label{fig:finetune}
\end{figure}

The success of large language models like GPT-4 \cite{achiam2023gpt} and Llama3 demonstrates the benefits of integrating diverse data sources during pre-training, enhancing their capabilities and generalizability across various real-world applications. This approach not only broadens the model's understanding of different data forms but also significantly boosts performance on specialized tasks through fine-tuning \cite{nguyen2024information} \cite{huang2024multi}. Inspired by these advancements, our work employs a cross-task data fusion strategy for LLM fine-tuning, aiming to enhance the model's effectiveness by combining insights from different financial tasks. Figure \ref{fig:finetune} illustrates the proposed fine-tuning method.

We curated and preprocessed a robust training set from two tasks: financial text classification and financial text summarization, to cover a wide range of real-world financial scenarios. We excluded the dataset for task 3, which focuses on texts related to three specific stocks, due to its narrow company-specific content. This selective integration forms the basis for fine-tuning a pre-trained LLM, equipping it to effectively understand and generate nuanced financial texts. After fine-tuning, we applied the enhanced model to each of the three tasks to evaluate its practical utility and performance across various financial applications.

\section{Experiment and Discussion}
In this section, we present technical details of our implementation and numerical results of our fine-tuned models on tasks 1, 2, and 3. We also compare the performance of these models on different tasks and present our observations on discrepancies between the two base models. 
\subsection{Experiment Setup}
 Mistral-7B and Llama3-8B are employed as the base LLM in this study. Due to the limit of computational resources, we perform fine-tuning using Low-Rank Adaptation (LoRA, \citet{hu2021lora}) with LoRA-$\alpha$ 16 and 4-bit quantization \cite{jacob2018quantization} to reduce the usage of GPU memory and to accelerate training. The models were trained and inferenced on two NVIDIA RTX-A6000 GPUs (each has 48GB DRAM) with one epoch. Our implementation employs PEFT, Quantization libraries and other pipelines provided in Huggingface\footnote{https://huggingface.co/}. 
We divided the training set portion of the validation set in the ratio of 80:20 for performance evaluation. The models are further tested and compared using the provided testing data sets.

\subsection{Experiment Results on Validation Set}
In preliminary experiments, we observed a significant difference in performance between the fine-tuned Mistral-7B and Llama3-8B models. Mistral-7B demonstrated superior predictive capabilities and produced well-formatted outputs that could be easily parsed to yield final predictions. In contrast, Llama3-8B required additional processing of its outputs through specific prompting, which could potentially alter the original outputs. Consequently, we decided to conduct all subsequent experiments using Mistral-7B.
\subsubsection{Task 1}
\begin{table}[h]
\centering
\begin{tabular}{|c|c|c|}
\hline
\textbf{Dataset}         & \textbf{ACC} & \textbf{F1}  \\ \hline
\textbf{No Fine-tune}          & 0.4997       & 0.1581       \\ \hline
\textbf{Task 1}          & 0.3490       & 0.3913       \\ \hline
\textbf{Task 1 + Task 2} & {\ul 0.6259} & {\ul 0.5634} \\ \hline
\end{tabular}
\caption{The performance for two models tasked with classifying sentences as either "premise" or "claim". It includes two key metrics: Accuracy (ACC) and F1 Score (F1). Model "Task 1" was fine-tuned using only the dataset from Task 1, while Model Task 1 + Task 2 used datasets from both Task 1 and Task 2 for fine-tuning.}
\label{tab:1}
\end{table}

Table~\ref{tab:1} illustrates that the fine-tuned LLMs have significantly improved reasoning for downstream-specific tasks. Furthermore, the LLMs, fine-tuned using the fused dataset, exhibit significant performance enhancements, where it achieves a 0.5634 F1 score. This evidence supports the notion that integrating different tasks can substantially enhance the reasoning capabilities of LLMs.

\subsubsection{Task 2}
\begin{table}[h]
\centering
\footnotesize
\begin{tabular}{|c|c|c|c|}
\hline
\textbf{Dataset}         & \textbf{Rouge-1} & \textbf{Rouge-2} & \multicolumn{1}{l|}{\textbf{BertScore}} \\ \hline
\textbf{Task 1}          & 0.4847           & 0.2921           & 0.6904                                  \\ \hline
\textbf{Task 1 + Task 2} & {\ul 0.4920}     & {\ul 0.3015}     & {\ul 0.6946}                            \\ \hline
\end{tabular}
\caption{The performance results for two models tasked with summarizing. It includes metrics for evaluating summarization: Rouge and Bert Score. Model "Task 1" was fine-tuned using only the dataset from Task 1, while Model Task 1 + Task 2 used datasets from both Task 1 and Task 2 for fine-tuning. }
\label{tab:2}
\end{table}

Table~\ref{tab:2} also demonstrates that the fine-tuned LLMs,by using the fused dataset, achieved significant performance gains in the text summarization task. This reinforces the idea that integrating various tasks can notably enhance the generalization capabilities of LLMs across different applications.

\subsubsection{Task 3}
We compare the three fine-tuned models based on Mistral-7B in Task 3. We exclude the models fine-tuned from Llama3-8B in this comparison because Llama3-based models cannot consistently produce trading decisions in the correct format. We fine-tuned three models:
\begin{enumerate}
    \item Model 1 is fined-tuned only using the training data from Task 1,
    \item Model 2 is fined-tuned only using the training data from Task 2,
    \item Model 3 is fined-tuned using the training data from Task 1 and Task 2.
\end{enumerate}
The three models are implemented in the FinMem framework as described in \citet{yu2024finmem} to generate trading decisions.
\begin{figure*}[h]
    \centering
    \begin{subfigure}{0.48\textwidth}
        \centering
        \includegraphics[width=\textwidth]{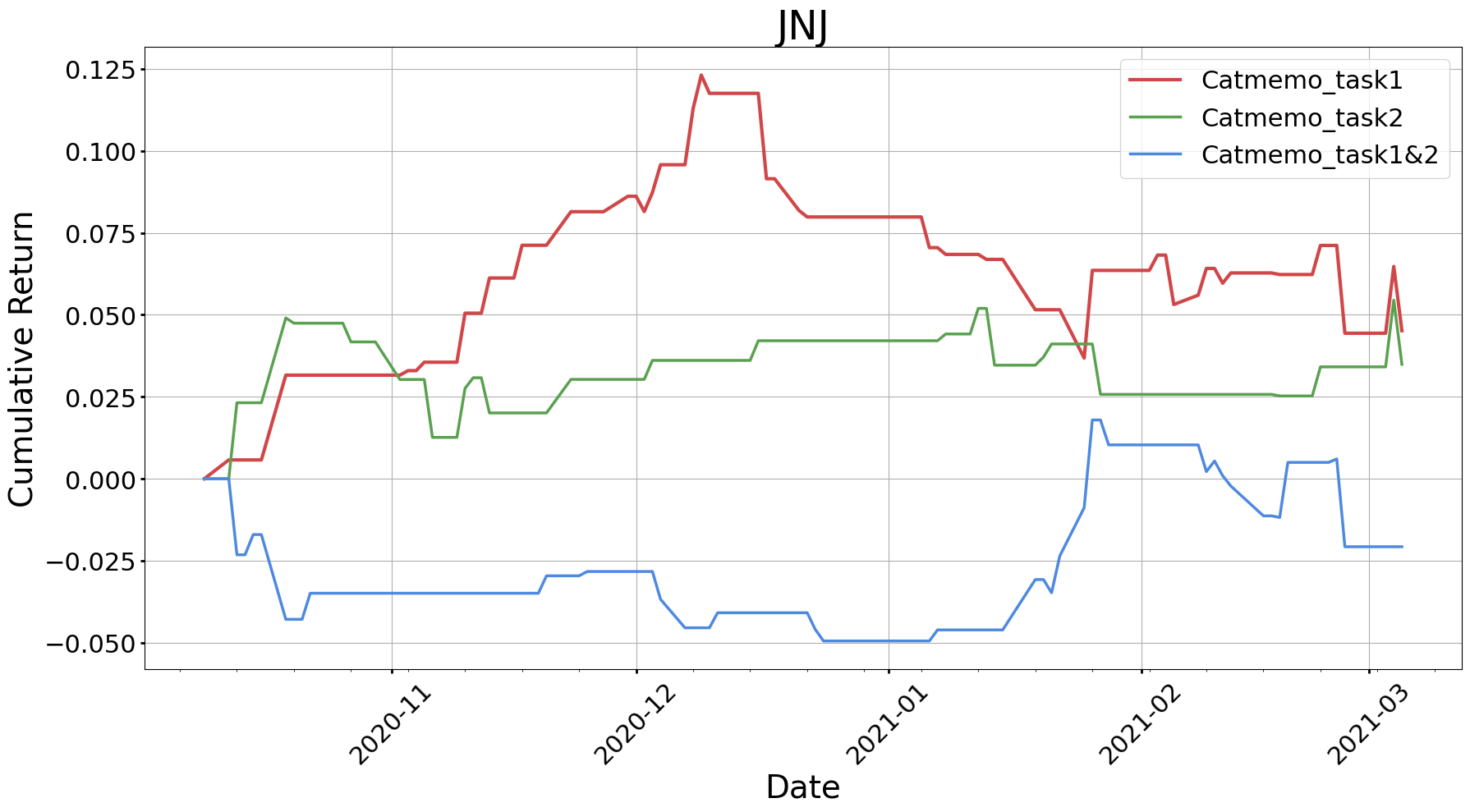}
        \caption{JNJ}
        \label{fig:JNJ}
    \end{subfigure}
    \hfill
    \begin{subfigure}{0.48\textwidth}
        \centering
        \includegraphics[width=\textwidth]{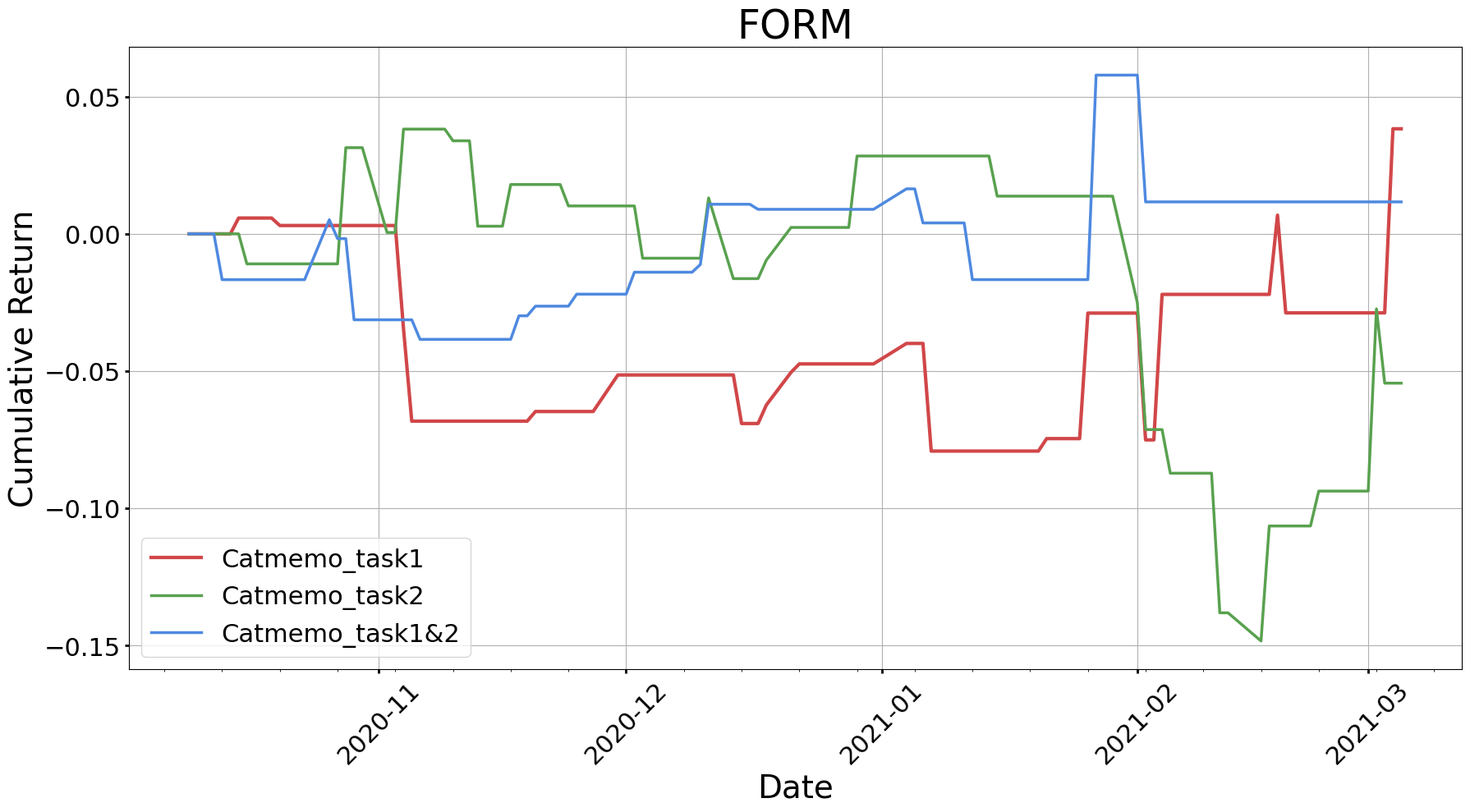}
        \caption{FORM}
        \label{fig:FORM}
    \end{subfigure}
    \vskip\baselineskip
    \begin{subfigure}{0.48\textwidth}
        \centering
        \includegraphics[width=\textwidth]{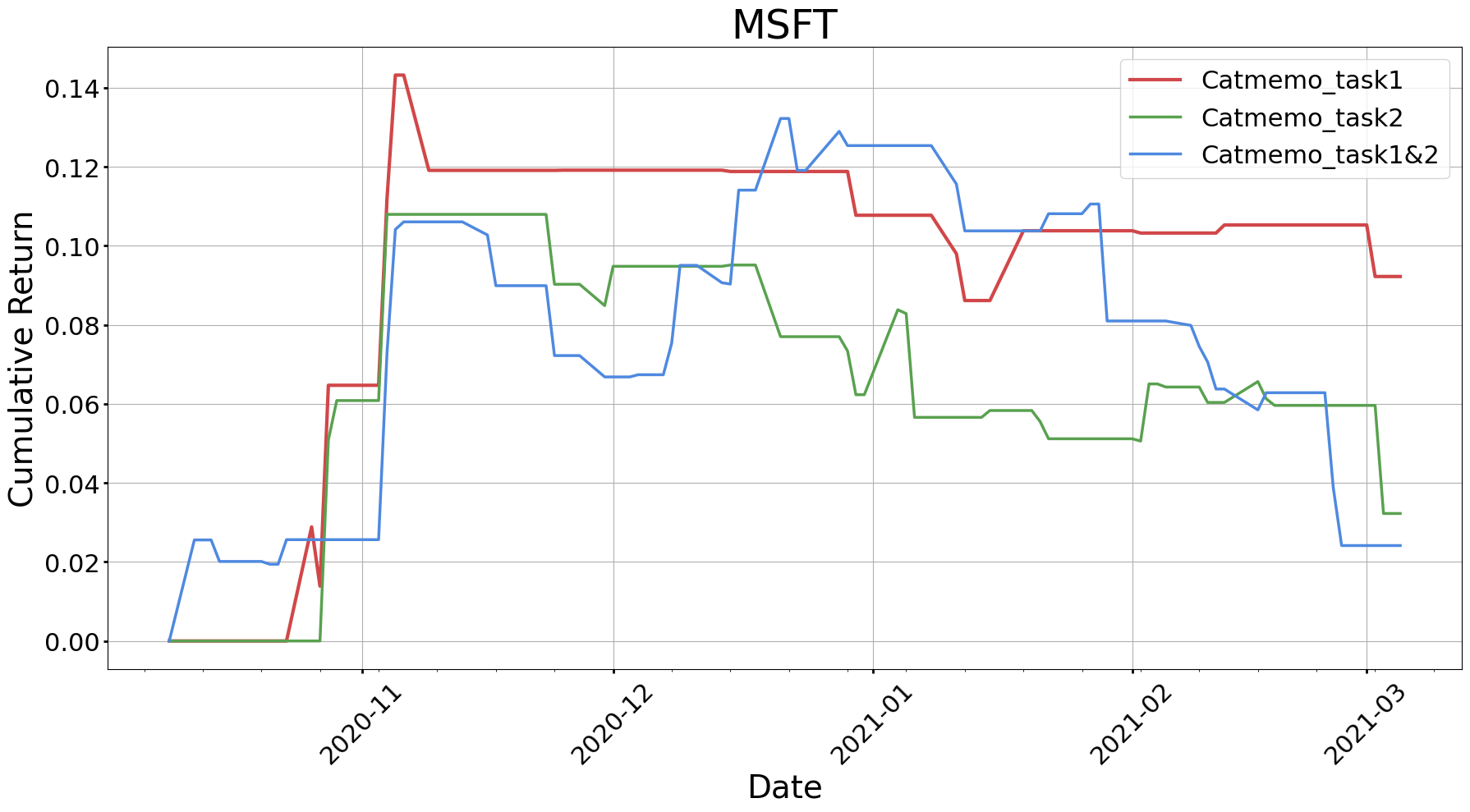}
        \caption{MSFT}
        \label{fig:MSFT}
    \end{subfigure}
    \hfill
    \begin{subfigure}{0.48\textwidth}
        \centering
        \includegraphics[width=\textwidth]{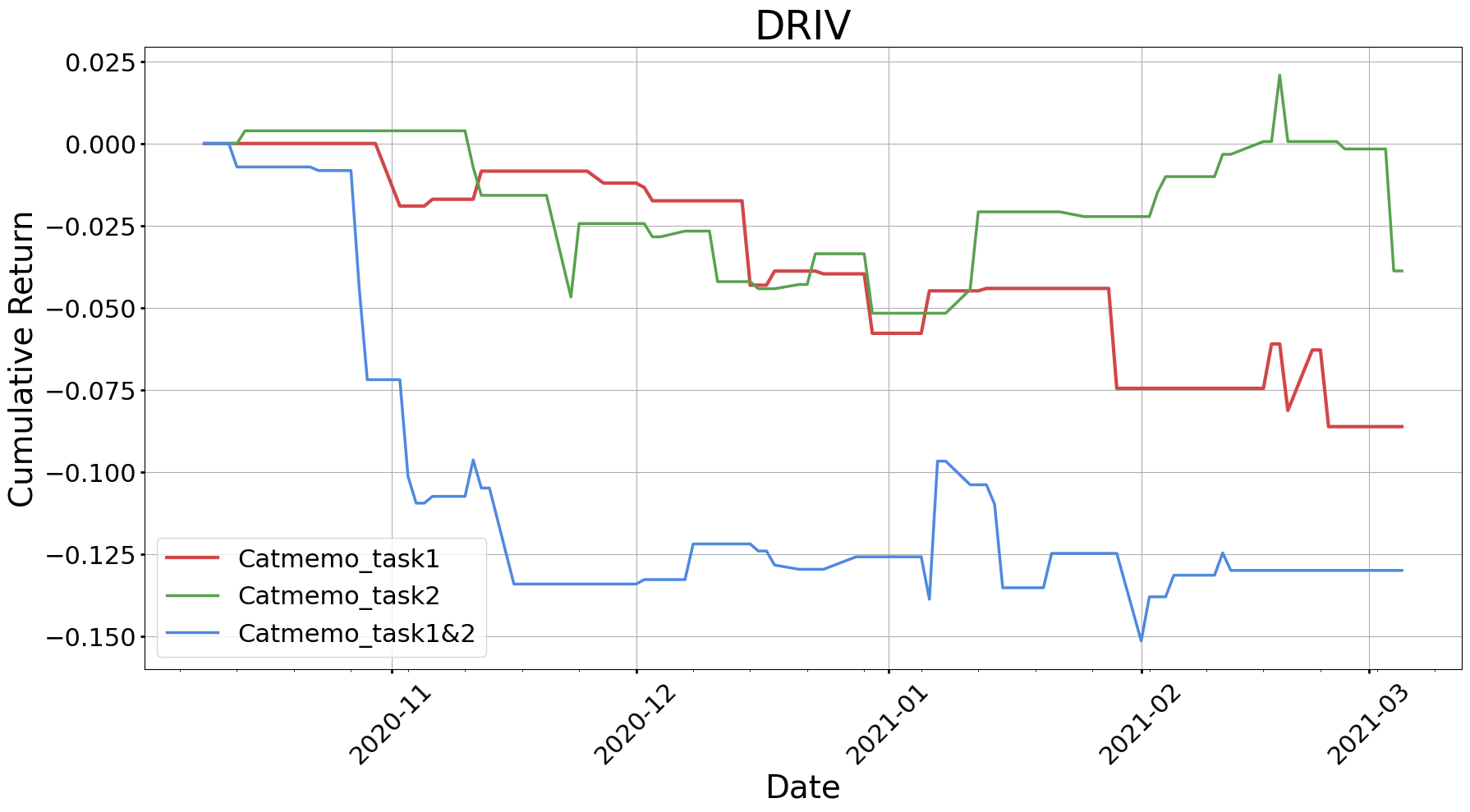}
        \caption{DRIV}
        \label{fig:DRIV}
    \end{subfigure}
    \caption{Comparison of Cumulative Returns in 4 Stocks}
    \label{fig:stock_comparison}
\end{figure*}

\begin{table*}[ht!]
\centering
\begin{adjustbox}{width=2\columnwidth,center}
\begin{tabular}{|c|c|c|c|c|c|c|c|c|c|c|c|c|}
\hline
\multirow{2}{*}{} & \multicolumn{3}{c|}{\textbf{FORM}} & \multicolumn{3}{c|}{\textbf{JNJ}} & \multicolumn{3}{c|}{\textbf{MSFT}} & \multicolumn{3}{c|}{\textbf{DRIV}} \\ \cline{2-13} 
                  & \textbf{Model 1} & \textbf{Model 2} & \textbf{Model 3} & \textbf{Model 1} & \textbf{Model 2} & \textbf{Model 3} & \textbf{Model 1} & \textbf{Model 2} & \textbf{Model 3} & \textbf{Model 1} & \textbf{Model 2} & \textbf{Model 3} \\
\textbf{CR} \tikz[baseline=-0.7ex]\draw[thick,->] (0,-0.18) -- (0,0.18);       & 0.038  & -0.054 & 0.012 & 0.045 & 0.035 & -0.021 & 0.092 & 0.032 & 0.024 & -0.086 & -0.039 & -0.130 \\ \hline
\textbf{SR}\ \tikz[baseline=-0.7ex]\draw[thick,->] (0,-0.18) -- (0,0.18);         & 0.440  & -0.574 & 0.176 & 0.927 & 0.898 & -0.506 & 1.594 & 0.564 & 0.418 & -2.139 & -0.834 & -2.291 \\ \hline
\textbf{SD}\ \tikz[baseline=-0.7ex]\draw[thick,->] (0,0.18) -- (0,-0.18);         & 0.014  & 0.015  & 0.010 & 0.006 & 0.006 & 0.009  & 0.009 & 0.009 & 0.009 & 0.006 & 0.007 & 0.009   \\ \hline
\textbf{AV}\ \tikz[baseline=-0.7ex]\draw[thick,->] (0,0.18) -- (0,-0.18);         & 0.217  & 0.237  & 0.165 & 0.101 & 0.097 & 0.102  & 0.144 & 0.143 & 0.144 & 0.101 & 0.116 & 0.142   \\ \hline
\textbf{MD}\ \tikz[baseline=-0.7ex]\draw[thick,->] (0,0.18) -- (0,-0.18);         & 0.084  & 0.175  & 0.046 & 0.084 & 0.059 & 0.144  & 0.056 & 0.074 & 0.104 & 0.084 & 0.059 & 0.144   \\ \hline
\end{tabular}
\end{adjustbox}
\caption{Performance Metrics Comparison Across Different Models and Datasets.}
\label{tab:task3_performance_comparison}
\end{table*}

We are interested in the performance discrepancies of these models trained on different datasets. Figure \ref{fig:stock_comparison} shows the return changes of the three models across four stocks during the testing period. Table \ref{tab:task3_performance_comparison} details the performance metrics of the models on different stocks. The models generate distinct strategies for all four assets, indicating sensitivity to the fine-tuning datasets. However, none of the models consistently produce profitable strategies. The Mistral-7B model is relatively small compared to state-of-the-art LLMs like OpenAI GPT-4 \cite{achiam2023gpt} and Google Gemini \cite{team2023gemini}, limiting its ability to solve complex tasks such as trading decisions. This aligns with the reported performance of other LLMs in \citet{xie2024finben}. Additionally, Model 3, trained on both datasets, does not outperform the models trained on each dataset individually. This could be due to the introduction of noise or conflicting information from combining datasets. Given that tasks 1 and 2 are not directly related to trading, it is reasonable that all three models perform poorly in this task.

\subsection{Experiment Results on Test Set}
Based on the above analysis, we selected the Mistral-7B model, fine-tuned through data fusion, for the final challenge testing. In Task 1, the model achieved an ACC of 0.711, an F1 score of 0.4199, and a Matthews correlation coefficient (MCC) of 0.6818. In Task 3, the integrated Sharp Ratio (SR) was -0.6199. These results are consistent with those observed in our validation set.

\section{Conclusion}
In this study, we fine-tuned LLMs using datasets that span multiple tasks, resulting in performance improvements in classification and summarization tasks. However, our approach did not yield positive results for the stock trading task. This outcome suggests that more complex financial tasks may require advanced data fusion steps. Furthermore, it underscores the need to explore the impact of incorporating larger datasets on the model’s performance after fine-tuning.

\section*{Limitation}
Our work relies on the pre-trained large language model at 7B/8B level with 4-bit quantization, we have not considered other parameter-level pre-trained models like Llama3-70B which will be explored in the future.


\bibliography{anthology}


\end{document}